\documentclass[superscriptaddress,nofootinbib, amsmath,amssymb,preprintnumbers,prdfloatfix,twocolumn,PRL]{revtex4-2}
\usepackage{CJK}
\usepackage{graphicx}
\usepackage{dcolumn}
\usepackage{upgreek}
\usepackage{slashed}
\usepackage{braket}
\usepackage{amsmath}
\usepackage{bm}
\usepackage[colorlinks=true,pdfstartview=FitV,breaklinks=true]{hyperref}
\usepackage[dvipsnames,table]{xcolor}
\hypersetup{urlcolor=BlueViolet,
	    citecolor=Plum,
	    linkcolor=PineGreen}
\usepackage{lipsum}	    
\usepackage{titlesec}
\usepackage{etoolbox}
\usepackage[normalem]{ulem}
\usepackage{subcaption}
 \usepackage{lettrine,Typocaps}

\usepackage{relsize}

\definecolor{offblue}{RGB}{23,80,153}

\newcommand{\ms}{m_{\ast}}
\newcommand{\mpsi}{m_{\psi}}
\newcommand{\mphi}{m_{\varphi}}
\newcommand{\vevphi}{\braket{\varphi}}
\newcommand{\mpsit}{\tilde{m}_{\psi}}
\newcommand{\mphit}{\tilde{m}_{\varphi}}
\newcommand{\ueff}{U_{\rm eff}}
\newcommand{\GeV}{\mathrm{GeV}}

\newcommand{\equaref}[1]{Eq.~(\ref{#1})}
\newcommand{\equasref}[2]{Eqs.~(\ref{#1})~and~(\ref{#2})}

\newcommand{\figref}[1]{Fig.~\ref{#1}}

\newcommand{\secref}[1]{Section~\ref{#1}}

\newcommand{\refref}[1]{Ref.~\cite{#1}}
\newcommand{\refsref}[2]{Refs.~\cite{#1}~and~\cite{#2}}

\usepackage[export]{adjustbox}

\begin{document}
\preprint{IPMU24-0042}

\title{\Large Black Holes from Fermi Ball Collapse}

 \author{Yifan Lu}
 \email{yifanlu@ucla.edu}
\affiliation{Department of Physics and Astronomy, University of California Los Angeles,\\ Los Angeles, California, 90095-1547, USA}

 \author{Zachary S. C. Picker}
 \email{zpicker@physics.ucla.edu}
\affiliation{Department of Physics and Astronomy, University of California Los Angeles,\\ Los Angeles, California, 90095-1547, USA}

  \author{Stefano Profumo}
 \email{profumo@ucsc.edu}
\affiliation{Santa Cruz Institute for Particle Physics, 1156 High St., Santa Cruz, CA 95064, USA}
\affiliation{Department of Physics, 1156 High St., University of California Santa Cruz, Santa Cruz, CA 95064, USA}

\author{Alexander Kusenko} 
 \email{kusenko@ucla.edu}
\affiliation{Department of Physics and Astronomy, University of California Los Angeles,\\ Los Angeles, California, 90095-1547, USA}
\affiliation{Kavli Institute for the Physics and Mathematics of the Universe (WPI), UTIAS \\The University of Tokyo, Kashiwa, Chiba 277-8583, Japan}

\begin{abstract}
\noindent Fermi balls are non-topological solitons that can naturally form in an early universe containing a dark sector with heavy fermions and an attractive interaction mediated by a light scalar field. We compute the Fermi ball mass and radius scaling relations when the potential of the scalar field $\varphi$ has a non-negligible quartic coupling $\lambda\varphi^4$. The resulting Fermi balls reach `saturation' very rapidly, even when their radius is much smaller than the effective Yukawa force range. These objects can therefore grow by mergers or by accretion of ambient dark fermions, until they become so dense that they fall within their Schwarzschild radius and collapse to black holes. This setup, therefore, provides an example of a rather natural and economical dark sector scenario for the formation of primordial black holes.
\end{abstract}

\maketitle

\section{Introduction}
\lettrine[lines=5,findent=3pt]{W}{hile} models with a small field content are sometimes considered more natural and appealing, complex new-physics sectors are often necessary and compelling: suffice to mention the benchmark QCD axion models~\cite{Dine:1981rt,Kim:1979if,Shifman:1979if,Zhitnitsky:1980tq},  which encompasses new Higgs fields or heavy quarks, or the lightest supersymmetric particle~\cite{Ellis:1983ew,Jungman:1995df}, which entails adding a significant amount of additional structure as well as new symmetries such as $R$-parity. Such rich dark sectors are not merely necessary appendages, but can bring compelling and interesting phenomenological consequences. For example, the asymmetric dark matter scenario~\cite{Nussinov:1985xr,Barr:1990ca,Kaplan:2009ag,Davoudiasl:2012uw,Petraki:2013wwa,Zurek:2013wia} includes a light dark scalar and a heavy dark fermion which allows it potentially to naturally explain why the dark matter and baryonic matter have similar abundances. 

There is therefore compelling motivation to carefully examine the cosmological and phenomenological consequences of interacting dark sectors, and specifically whether macroscopic dark structures may readily form as a result. The stability and evolution of such macroscopic objects is similarly important. For a number of reasons which we detail here and will further elaborate upon in a forthcoming study, of particular interest is the question of whether primordial black holes~\cite{pbh,Hawking:1971ei,Carr:1974nx,Chapline:1975ojl,Green:2016xgy,Cotner:2016cvr,Cotner:2019ykd,Flores:2020drq,Carr:2020gox} could result from such interacting, rich dark sectors.

An extensive literature exists on the formation of macroscopic objects in the dark sector. One of the simplest dark sectors realizations consists of a species of fermions $\psi$ and (usually) a scalar $\varphi$ that mediates the interaction between fermions, and there have been many predictions of the formation of macroscopic objects in this scenario. We will focus here on the formation of compact objects of these fermions called \textit{Fermi balls}, non-topological solitons roughly analogous to neutron stars but entirely in the dark sector. These can be produced from nucleosynthesis of asymmetric dark matter \cite{Kaplan:2009ag, Hardy:2014mqa, Wise:2014jva, Wise:2014ola, Gresham:2017cvl}, trapped fermions from a first-order phase transition in the dark sector \cite{Kawana:2021tde, Huang:2022him, Kawana:2022lba, Lu:2022jnp}, dissipative dark matter in the early matter dominated era \cite{Chang:2018bgx, Bramante:2024pyc}, or early structure formation induced by a long range force \cite{Gradwohl:1992ue, Gubser:2004uh, Nusser:2004qu, Amendola:2017xhl, Savastano:2019zpr, Flores:2020drq, Flores:2023zpf, Domenech:2021uyx}, to mention a few well-motivated possibilities. 

If the Fermi balls form sufficiently early and are stable, they would themselves comprise a viable cold dark matter candidate, subject of course to the ensemble of observational and experimental constraints on generic macroscopic dark matter candidates. While this could be considered a catastrophic scenario to direct detection experiments, it may be very interesting for astrophysical searches and observations. 

However, the stability of the resulting compact object, especially when a Yukawa interaction is introduced between the dark fermion and the scalar, can be rather involved: in particular, the question of whether the pressure-supported Fermi ball can continue to collapse to a black hole crucially depends on the behavior of the fermion-scalar system; this issue was originally stated and studied in \refsref{Lee:1986tr}{Lee:1991ax} as non-topological fermion `soliton stars'. The non-trivial scaling behavior of the Fermi ball radius and mass was further studied in Refs.~\cite{Gresham:2017zqi, Xie:2024mxr, DelGrosso:2023dmv, DelGrosso:2023trq, DelGrosso:2024wmy}. 

In this work we investigate the Fermi balls' {\it equations of state} and {\it stability} in the scenario where the scalar self-coupling is non-negligible. In particular, the existence of a $\lambda \varphi^4$ term in the scalar potential, an inclusion which is in any case required for renormalizability, dramatically modifies the interior behavior of the Fermi ball, allowing them to reach `saturation'---the state where the effective Yukawa interaction length \textit{inside} Fermi ball is comparable to the size of ball, and equivalent to the use of the term for ordinary neutron stars---at significantly smaller fermion particle numbers. 

Reaching saturation has two important consequences: Firstly, the Yukawa force range \textit{outside} the Fermi ball can be much larger than the saturated ball radius, allowing for long-range interactions which we derive here; Secondly, the Fermi ball mass and radius relations during saturation imply that {\it heavier balls become increasingly compact}, so that they will naturally eventually have radius smaller than their Schwarzschild radius and collapse to a black hole. Without the $\varphi^4$ potential, these two behaviors are not simultaneously compatible---a Fermi ball below saturation would have mass and radius which grow proportionately, but after saturation it no longer experiences long-range forces. 

We then have a natural black hole formation mechanism---more akin to neutron stars which exceed the Tolman--Oppenheimer--Volkoff limit if their mass becomes too large---along with long-range forces which actually allow the Fermi balls to continue growing until they reach collapse. In previous work (see e.g.~\cite{Lu:2022jnp}), the possibility of black hole formation was considered to be only sensitive to the initial conditions of the halos and of the model parameters, since a Fermi ball which could \textit{continue} growing would \textit{not} eventually collapse to a black hole---either a collapsing halo was always destined to be a black hole, or not. We then put such black hole formation on more solid footing here, showing explicitly that degeneracy pressure will be overcome when the Fermi balls become sufficiently dense.

The structure of the remainder of this study is as follows: In Sec.~\ref{sec:fballs} we review the use of mean-field theory for determining the mass and radius of a Fermi ball, and extend this analysis to the case where the scalar field has a $\varphi^4$ potential. In Sec.~\ref{sec:interaction} we estimate the long range force between Fermi balls in this scenario, before discussing their formation and other potential cosmological consequences in Sec.~\ref{sec:cosmology}. We provide a discussion and summary of our results, and draw our conclusions, in the final Sec.~\ref{sec:conclusions}.

\section{Fermi ball stability}\label{sec:fballs}

\noindent We begin with a review of the Fermi ball equations of state in the scenario where the scalar field has {\it no potential}, i.e. no self-interactions, following previous work in \refref{Gresham:2017zqi, Xie:2024mxr, DelGrosso:2023dmv, DelGrosso:2023trq, DelGrosso:2024wmy}. We consider  a particle model consisting of a heavy dark fermion $\psi$ coupled to a light dark scalar $\varphi$, with Lagrangian,
\begin{equation}
    \mathcal{L}=\bar{\psi}\left(i \slashed{\partial}-\left(m_\psi-y \varphi\right)\right) \psi+\frac{1}{2}(\partial \varphi)^2-\frac{1}{2} m_{\varphi}^2 \varphi^2-V(\varphi)~.
\end{equation}
For simplicity, we will assume that these interactions take place entirely in the dark sector with no coupling to the Standard Model---we note, however, that one may also consider our Lagrangian as a demonstrative toy model which could be found within a more complicated, beyond the Standard Model sector. 

The presence of the Yukawa coupling produces an attractive interaction between fermions and allows for the emission of scalar radiation from accelerated fermions, making them effectively {\it dissipative}. We discuss Fermi ball formation in the early universe later; we assume here to have a dense gas of $N$ fermions, and we are interested in understanding when there is sufficient pressure to resist collapse. This will in turn determine the Fermi ball properties and the possibility of black hole formation. 

When the scalar acquires a vacuum expectation value (VEV), the Yukawa term can also be treated as a shift in the effective fermion mass,
\begin{equation}
    \ms = \mpsi - y \vevphi.
\end{equation}
This effect is crucial in the Fermi ball many-body system, as the large fermion density induces a non-zero scalar VEV to minimize the energy. Notably, in the absence of $V(\varphi)$, the reduced fermion mass quickly leads to relativistic constituents and a large scalar VEV as the fermion number $N$ grows \cite{Gresham:2017zqi}, and the non-relativistic Yukawa energy breaks down in this regime. To proceed, we adopt the formalism of relativistic mean field theory using the Walecka model \cite{Walecka:1974qa, Serot:1991st}. At high scalar occupancy number, the scalar field can be replaced by its classical VEV, while the fermions are described using the Thomas-Fermi approximation \cite{Spruch:1991zz}. The profile of a spherically symmetric Fermi ball is then determined by the classical field $\varphi(r)$ and by the Fermi momentum $k_F(r)$, with  total energy given by
\begin{align}
    E\left[\varphi(r), k_F(r)\right]=\int_0^R & dr \, 4 \pi r^2\Bigg[\frac{1}{2}(\nabla \varphi)^2+\frac{1}{2} m_\varphi^2 \varphi^2 + V(\varphi) \nonumber\\
    &+\frac{g_{f}}{2 \pi^2} \int_0^{k_F(r)} dk\, k^2 \varepsilon(k, \varphi)\Bigg].
    \label{energy}
\end{align}
In the expression above $g_f = 2$ is the degeneracy of a Dirac fermion, and we assumed zero temperature. The shift in the fermion effective mass is included in the fermion energy, $$\varepsilon(k, \varphi) = \sqrt{k^2 + (\mpsi - y \varphi)^2}.$$

The fermion number $N$ is a conserved charge due to the global $U(1)$ symmetry of the Lagrangian. At zero temperature, it is related to the Fermi momentum profile by
\begin{equation}
    N=\frac{2 g_{f}}{3\pi} \int_0^R dr\, r^2 k_F^3(r).
    \label{fermionnumber}
\end{equation}
Conservation of fermion number implies that one needs to minimize the energy functional with an additional Lagrange multiplier:
\begin{align}\label{variation}
    \delta E - \mu \delta N&=\left(\frac{\delta E}{\delta \varphi}-\nabla \frac{\delta E}{\delta \nabla \varphi}\right) \delta \varphi\\ &+\left(\frac{\delta E}{\delta k_F}-\mu \frac{\delta N}{\partial k_F}\right) \delta k_F\nonumber
    =0.
\end{align}
The variation of $\varphi$ is equivalent to a classical field equation, giving the differential equation for the scalar profile,
\begin{equation}
    \frac{1}{r} \frac{d^2 (r\varphi)}{dr^2} = m_{\varphi}^2 \varphi + \frac{\partial V}{\partial \varphi} - y\braket{\bar{\psi} \psi},
    \label{phiode}
\end{equation}
where the fermion scalar density $\braket{\bar{\psi} \psi}$ is given by,
\begin{equation}
    \braket{\bar{\psi} \psi} = \frac{g_{f}}{2 \pi^2} \int_0^{k_F(r)} dk\, \frac{k^2 \ms}{\sqrt{k^2+\ms^2}}~.
\end{equation}
The Lagrange multiplier $\mu$ can be viewed  physically as the chemical potential of the system, and it can be derived using the variation of $k_F$ from \equaref{variation}:
\begin{equation}
    \mu = \sqrt{k_F(r)^2+\left(m_\psi-y \varphi(r)\right)^2}.
    \label{mu}
\end{equation}
It is worth noting that $\mu$ is spatially constant and so independent of $r$, allowing us to use \equaref{mu} when expressing $k_F$ in terms of $\varphi$. Then the Fermi ball profile can be solved entirely from \equaref{phiode}. However, this differential equation is a boundary value problem, rather than a more straightforward initial value problem. Specificially we have the conditions,
\begin{equation}
   \frac{d \varphi}{d r}\Bigr |_{r=0}=0, \quad \varphi(+\infty) = \varphi_0,
   \label{eq:bdry}
\end{equation}
where $\varphi_0$ is the value of $\varphi$ at the true vacuum. Once the scalar field profile is obtained from \equasref{phiode}{eq:bdry}, the Fermi ball radius can be determined by finding the position where $k_F(R) = 0$ (or $\ms(R) = \mu$). 

It is convenient to define the dimensionless quantities,
\begin{align}
    &\tau = \mu r, \quad \rho(\tau) = \frac{\mpsi - y \varphi(\tau)}{\mu},\nonumber\\ &\mpsit = \frac{\mpsi}{\mu}, \quad \mphit = \frac{\mphi}{\mu}.
\end{align}
For now, we set $V(\varphi)=0$ to demonstrate the general behavior of the Fermi ball and its importance will be highlighted in the following section. We can recast \equaref{phiode} and the boundary conditions \equaref{eq:bdry} in terms of the effective fermion mass $\rho(\tau)$ and effective potentials, where derivatives are taken with respect to $\tau$:
\begin{align}
    &\rho^{\prime \prime}=-\frac{2}{\tau} \rho^{\prime}-\frac{\partial U_{f}}{\partial \rho}+\frac{\partial U_{s}}{\partial \rho}, \nonumber\\
    &\frac{d \rho}{d \tau}\Bigr |_{\tau=0}=0,\quad  \rho(+\infty) = \mpsit.
    \label{rhoode}
\end{align}
In this equation, the fermion scalar density is integrated and its contribution to the effective potential is given by
\begin{equation}
    U_f = \frac{y^2 g_{f}}{48 \pi^2}\left(3 \rho^4 \log \frac{1+\sqrt{1-\rho^2}}{\rho}-\sqrt{1-\rho^2}\left(5 \rho^2-2\right)\right)~,
\end{equation}
for $0<\rho<1$ and where the upper bound of $\rho$ is achieved at the boundary of the Fermi ball and $U_f = 0$ outside the ball (when $\rho > 1$), and the scalar effective potential reads
\begin{equation}
    U_s = \frac{1}{2}\mphit^2\left(\rho-\mpsit\right)^2.
\end{equation}
The input data for solving the Fermi ball profile includes parameters in the Lagrangian, together with a choice of the chemical potential $\mu$. Given a solution to \equaref{rhoode}, the fermion number of the ball can be calculated from \equaref{fermionnumber} while the radius is found by the root of $\rho(\tau) = 1$.

\subsection{Scalar field equation of motion}

\begin{figure*}[ht!]
    \begin{subfigure}{0.48\textwidth}
        \includegraphics[width=\textwidth]{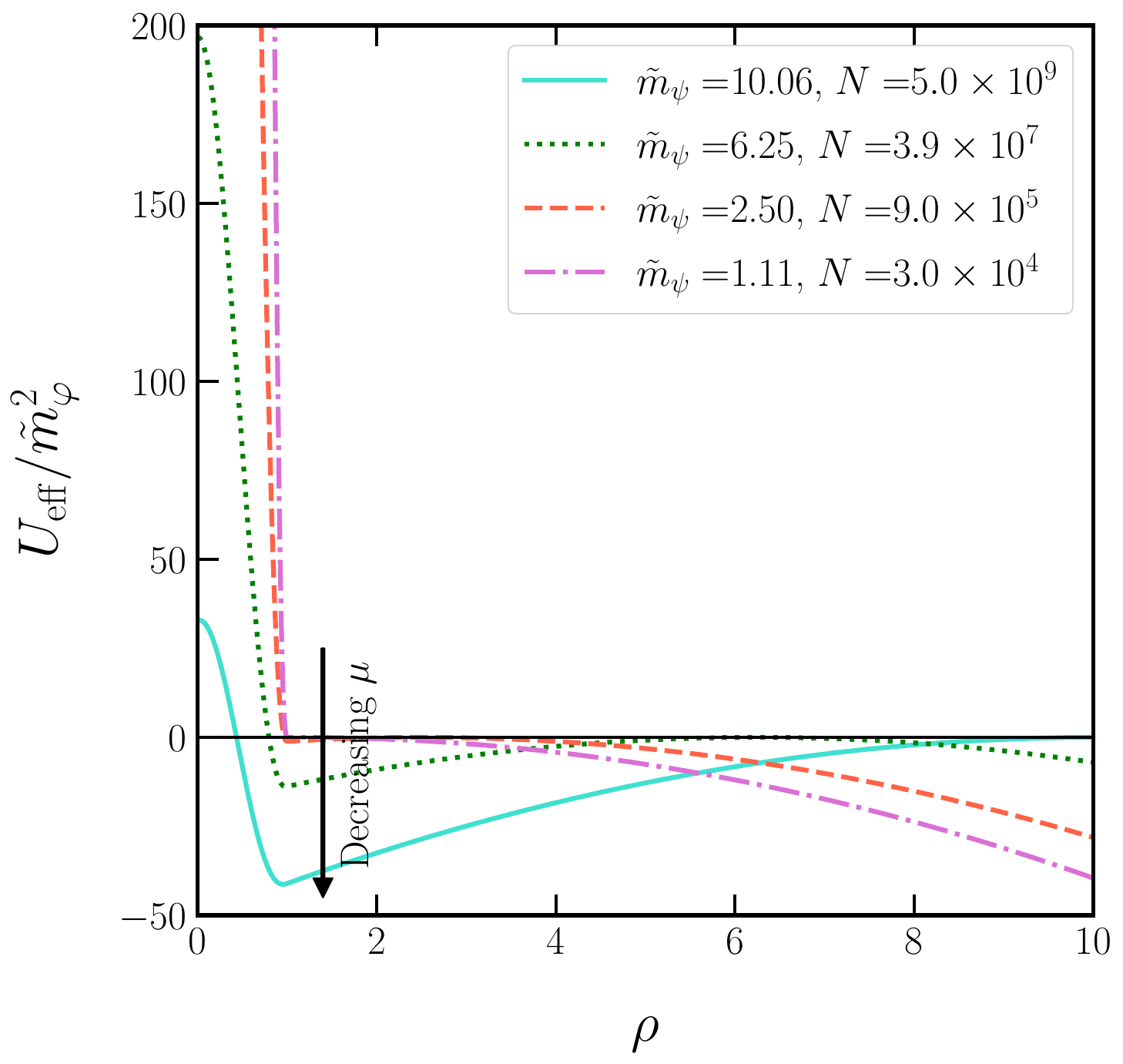}
    \end{subfigure}
    \begin{subfigure}{0.48\textwidth}
        \includegraphics[width=\textwidth]{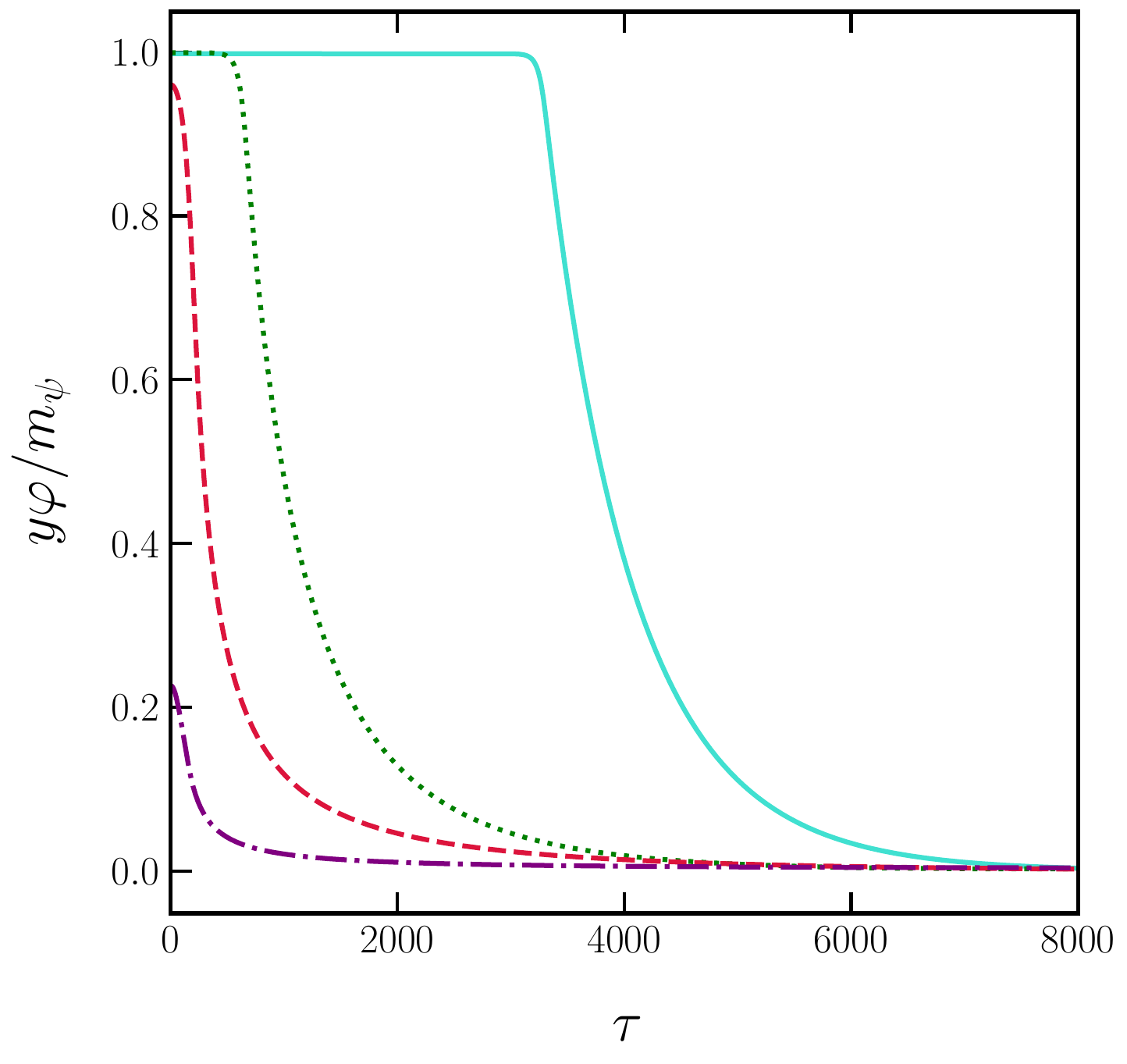}
    \end{subfigure}
    \caption{The shape of the effective potential and the scalar field profile are demonstrated for several choices of $\mu$ with $V(\varphi)=0$. Line colors in both panels correspond to the same $\mu$. The fermion and scalar masses are set to $\mpsi/\mphi = 10^4$ and $y=0.1$.}
    \label{fig:profiles}
\end{figure*}

\noindent In general, there is no analytical solution to \equaref{rhoode} and it cannot be straightforwardly integrated numerically due to the two-sided boundary conditions. Before discussing numerical procedures for solving this issue, it is instructive to study the qualitative behavior of the solution to gain some intuition. 

As pointed out in \refref{Xie:2024mxr}, the differential equation for the (dimensionless) effective mass shares great resemblance to the scalar field equation of motion in the context of Q-balls \cite{Coleman:1985ki}. The properties of the solution can be better understood using analogies in classical mechanics, where $\rho(\tau)$ can be thought of as the position of a ball at time $\tau$, sliding in an effective potential $\ueff = U_f - U_s$. The first term on the right hand side of \equaref{rhoode} introduces an additional time dependent friction force. With a proper choice of $\mu$, the effective potential has the structure of two hilltops, where the \textit{left hilltop} at $\rho<1$ is formed from the contribution of $U_f$ and the \textit{right hilltop} at $\rho>1$ always has height 0 due to $U_s$. Examples of $\ueff$ with varying $\mu$ are shown in \figref{fig:profiles}. 

The boundary conditions require a proper choice of the initial position $\rho_i$ so that the ball's motion starts at $\rho < 1$ with zero initial velocity, and exactly stops at the peak of the right hilltop when $\tau \rightarrow \infty$. When the left hilltop is higher than the right hilltop, such solution always exists and this can be proved using the ball-shooting arguments in \refref{Coleman:1985ki}: we can choose $\rho_i$ so that it is arbitrarily close to the left hilltop, then the ball will stay there for long enough time so that the friction force becomes negligible. Then the ball's energy is approximately conserved through the motion and it ensures the ball will always reach---but not necessarily stop---at the right hilltop (`overshooting'). Next, we can shift $\rho_i$ to the right so that its terminal velocity at the right hilltop decreases to zero terminal velocity. Such position always exists, as can be seen by considering $\rho_i < 1$ at the zero of $\ueff$: due to the non-conservative friction force, the ball will never be able to reach the right hilltop (`undershooting') and by scanning $\rho_i$ continuously from the overshooting scenario to the undershooting scenario, a proper choice of $\rho_i$ can be found.

From the arguments above, we see that the existence of the Fermi ball crucially depends on the existence and height of the left hilltop, which is, in turn, determined by the choice of $\mu$; as $\mu$ decreases, the height of the left hilltop decreases due to the change in $U_s$ (see \figref{fig:profiles}). Eventually, at some $\mu_{\rm min}$, the value of $\ueff$ at the left hilltop  becomes negative, preventing the formation of Fermi balls. On the other hand, the stability of the Fermi ball requires $\mu < \mpsi$. Therefore, a proper choice of $\mu$ must satisfy $\mu_{\rm min} < \mu < \mpsi$. 

In summary, given a value $\mu<\mpsi$, the existence of a Fermi ball numerical solution requires the following: (1) the existence of the left hilltop at $\rho<1$ (since the right hilltop always exists), and (2) that the height of the left hilltop is positive. Once these two requirements are satisfied, a scan of $\rho_i$ between the position of the left hilltop and the left zero of $\ueff$ can be performed in order to find the proper initial condition $\rho_0$. The Fermi ball profile is finally determined by solving \equaref{rhoode}, again using initial $\rho_0$. Solutions of the profile with varying $\mu$ are shown in \figref{fig:profiles}.

The fermion number $N$ is also closely related to $\mu$. The general relation between $N$ and $\mu$ can be understood from the shape of the effective potential. As $\mu$ decreases, the height of the left hilltop approaches the height of the right hilltop. This forces $\rho_0$ to get closer to the left hilltop and, consequently, it takes longer `time' for the ball to reach $\rho=1$, the boundary of the Fermi ball. Thus the Fermi ball's radius (related to the time in the heuristic ball shooting argument) and the fermion number increase with decreasing $\mu$. As we see in \figref{fig:profiles}, for small enough $\mu$ (or large $N$), the Fermi ball profile becomes flat due to the similar heights of the hilltops in $\ueff$. The Fermi ball reaches the saturation regime, similar to nuclear saturation in the Walecka model \cite{Walecka:1974qa, Serot:1991st}. Interestingly, analytic solutions in the saturation limit do exist, allowing us to obtain closed form formulas for the Fermi ball radius and fermion number.

\subsection{Fermi ball saturation and scaling}
\noindent There are three important regimes for the Fermi ball behavior: The first regime occurs at small $N$, so that $\varphi$ is small and the  fermions $\psi$ are non-relativistic. This regime is interesting for dark matter bound state formation \cite{Gresham:2017cvl, Gresham:2017zqi} and we refer to this regime as the {\it `nugget' regime}. Once $N$ is sufficiently large, one enters {\it `sub-saturation'} Fermi balls. This regime is a transition between the nugget Fermi ball and the saturated Fermi ball, and it occurs whenever the Fermi ball radius is smaller than the effective scalar mass inside the Fermi ball, $r<1/m_{\varphi,\rm eff}$. Finally, we have the {\it `saturation'} regime, where the effective scalar range is smaller than the Fermi ball radius and the Fermi ball profile becomes flat. We will not discuss the nugget regime here, but we should carefully analyze the qualitative difference in behavior between the sub-saturation and saturation regimes, since the difference in behavior here is a driver of the present analysis.\\

\noindent \textbf{Saturation regime:} Motivated by the flat profile of the Fermi ball at large $N$, we can solve for $\varphi$ and $k_F$ by simply assuming them to be constant and by directly minimizing the total energy without solving the equation of motion. In this sense, the saturated Fermi ball is analogous to the thin-wall regime of Q-balls. The energy \equaref{energy} in the saturation limit receives contributions from the bulk and surface, where the former scales as $N$. The surface contribution arises due to the change in the scalar VEV near the boundary of the Fermi ball, and generally scales as $N^{2/3}$. At large $N$, the surface term can be neglected at  leading order, and the energy simplifies to,
\begin{align}
    E=\frac{4 \pi}{3} R^3&\Bigg(\frac{1}{2} m_{\varphi}^2 \varphi^2+V(\varphi)+\nonumber\\
    &+\frac{g_{f}}{2 \pi^2} \int_{0}^{k_F} d k\, k^2 \sqrt{k^2+\ms^2}\Bigg)~,
\end{align}
where we restored the additional scalar potential $V(\varphi)$. Then the Fermi ball can be solved by,
\begin{equation}
    \frac{\partial E}{\partial \varphi}=0 \quad \text { and } \quad \frac{\partial E}{\partial k_F}=0~,
    \label{minimize}
\end{equation}
leading to algebraic equations for $\varphi$ and $k_F$. Fermion number conservation follows from imposing
\begin{equation}
    N=\frac{2 g_{f}}{9 \pi} R^3 k_F^3,
\end{equation}
which also determines $R$ for a given $N$. 

As an example, we work out the solution for $V=0$ which has been considered in \refsref{Gresham:2017zqi}{Xie:2024mxr}. The total energy contains cubic and quartic terms of $\ms$ from the Fermi momentum integration. In the absence of $V(\varphi)$ and $\mphi \ll \mpsi$, the energy is minimized when $\ms\rightarrow 0$ or $y \varphi \rightarrow \mpsi$. This can also be understood loosely using a pressure balancing argument: the Yukawa attraction requires a high degeneracy pressure to balance, leading to highly relativistic and approximately massless fermions. Taking $\ms \ll k_F$, \equaref{minimize} gives
\begin{equation}
\begin{aligned}
& m_{\varphi}^2 \varphi=\frac{g_{f}}{4 \pi^2} k_F^2 \ms, \\
& \mphi^2 \varphi^2=\frac{g_{f}}{12 \pi^2} k_F^4.
\end{aligned}
\end{equation}
The above equations can be easily solved, yielding
\begin{equation}
\begin{aligned}
& m_*=\left(\frac{4 \pi^2}{3 y^2 g_{f} }\right)^{1/2} \mphi, \\
& k_F=\left(\frac{12 \pi^2}{y^2 g_{f}}\right)^{1/4}\left(\mphi m_\psi \right)^{1/2}, \\
& R = \left(\frac{243}{1024 \pi^2 g_f}\right)^{1/12}\left(\frac{y}{\mphi m_\psi}\right)^{1/2} N^{1/3}.
\end{aligned}
\end{equation}
Notably, the solution for $\ms$ is consistent with our ansatz and the Fermi ball radius scales geometrically with $N^{1/3}$. Because the radius only scales with $N^{1/3}$ for saturated Fermi balls, there is a well-defined limit where the ball becomes so massive that it enters its own Schwarzschild radius $R_s\equiv 2M/M^2_\mathrm{Pl.}$, and collapses to a black hole. This was discussed in Ref~\cite{Flores:2023zpf} as a potential mechanism for primordial black hole formation.\\


\noindent \textbf{Sub-saturation regime:} The alternative is to consider Fermi balls with $R \mphi \ll 1$. The mass and radius scaling with the number of fermions $N$ is radically different in this scenario, as pointed out in Ref.~\cite{Gresham:2017zqi}:
\begin{align}
    R \sim N^{2/3}, \quad M \sim N^{2/3}~.
\end{align}
This mass scaling relation indicates an increased binding energy per constituent with increasing $N$, which is a feature of unsaturated Fermi balls. In this case the ball does not collapse to a black hole, because  the mass and the radius increase with $N$ in the same way. A Fermi ball growing by either accreting more free fermions or by merging with other Fermi balls will simply increase in size until it reaches saturation. 

However, sub-saturated Fermi balls have a unique feature: {\it the ball radius is much smaller than the Yukawa interaction length}, meaning that long-range Fermi ball -- Fermi ball interactions are possible. In contrast,  saturated Fermi balls only interact with other Fermi balls when they are as close as almost the sum of their radii. In the context of forming black holes, this presents a trade-off: In the regime where the Fermi balls interact with one other, they cannot collapse to black holes, and in the regime where they can collapse to black holes, they are not strongly interacting. 

We emphasize that {\it  all these properties rely on our assumption that the scalar potential is negligible}. In the next section, we  rederive the Fermi ball behavior in the scenario where the potential is not negligible and show that this behavior is altered significantly, leading to black hole production.



\subsection{Fermi balls in $\varphi^4$ theories}
\label{longrangeFB}


\noindent Let us  consider Fermi balls with a potential $V(\varphi) = \lambda \varphi^4/4$, as would in fact be required for renormalizability. If the quartic interaction  is absent at tree level, it can be generated via a fermion loop, and the required counterterm would not be available in the absence of a tree-level coupling.  Thus the usual requirement that a renormalizable lagrangian must include all dimension-four terms consistent with the symmetries forces us to include the $\varphi^4$ term.  Of course, the loop corrections modify the mass of the scalar field and, in some range of parameters, may require fine-tuning of the (non-physical) bare mass in the Largangian to yield a small physical scalar mass.  
Our results are readily generalized to other types of scalar potentials. 

A similar energy minimization procedure yields the following transcendental equations:
\begin{align} \label{phi4equation}
    &\mphi^2 \varphi+\lambda \varphi^3 = \frac{y g_{f}}{4 \pi^2}\Bigg( \ms^3 \log \frac{\ms}{k_F+\sqrt{k_F^2+\ms^2}}+\nonumber\\
    &\hspace{3.3cm} +k_F m_* \sqrt{k_F^2+m_*^2}\Bigg)~, \\
    &\frac{1}{2} \mphi^2 \varphi^2+\frac{1}{4} \lambda \varphi^4 = \frac{g_{f}}{16 \pi^2}\Bigg(\sqrt{k_F^2+\ms^2}\left(\frac{2}{3} k_F^3 - k_F \ms^2\right) + \nonumber\\
    &\hspace{3.3cm}-\ms^4 \log \frac{\ms}{k_F+\sqrt{k_F^2+\ms^2}}\Bigg).
\end{align}
The above equations have interesting analytic solutions in two branches. In the weak coupling branch ($\lambda \ll (\mphi / y \mpsi)^2$), the additional scalar potential only affects the solution perturbatively, and the Fermi ball still saturates with relativistic fermions, as shown in \refref{Gresham:2017zqi}. 

There exists, however, a second branch of solutions with $1 \gtrsim \lambda \gg (\mphi / y \mpsi)^2$ and small $\mphi$ which we find to have drastically different behavior; we refer to this class of solutions  as the {\it`strong coupling branch'}, which, we show below will lead to significant Fermi ball -- Fermi ball long range interaction. In this regime we can neglect the scalar mass contribution and the scalar field energy is dominated by $V(\varphi)$. Specifically this occurs when,
\begin{align}
    \mphi \ll \lambda^{1/2} \alpha^{2/3} \mpsi / y~,
\end{align}
where $\alpha\equiv (g_f y^4)/(6\pi^2\lambda)$. Physically, the $\varphi^4$ interaction provides a {\it repulsive force} in addition to the original Fermi degeneracy pressure that helps balance the attractive Yukawa force. Based on this intuition, as $y$ decreases, $\varphi$ inside the Fermi ball should also decrease. Therefore, for sufficiently small $y$ (defined carefully below), the strong coupling branch features non-relativistic fermions and a small scalar VEV. 

In the $k_F \ll \ms$ limit, expanding \equaref{phi4equation} gives,
\begin{equation}
\begin{aligned}
\tilde{\varphi}^3 &= \alpha \kappa^3, \\
\frac{5}{4} \tilde{\varphi}^4 &= \alpha \kappa^5 \frac{\mpsi}{\ms},
\end{aligned}
\end{equation}
where we defined the dimensionless variables, 
\begin{equation}\label{eq:dims}
    \tilde{\varphi} = \frac{y \varphi}{\mpsi}, \quad \kappa = \frac{k_F}{\mpsi}~.
\end{equation}
Using our ansatz $\ms \rightarrow \mpsi$, we obtain the solutions,
\begin{equation}
    \kappa = \frac{5}{4} \alpha^{1/3}, \quad \tilde{\varphi} = \frac{5}{4}\alpha^{2/3}~.
\end{equation}

\begin{figure*}[ht!]
    \begin{subfigure}{0.48\textwidth}
        \includegraphics[width=\textwidth]{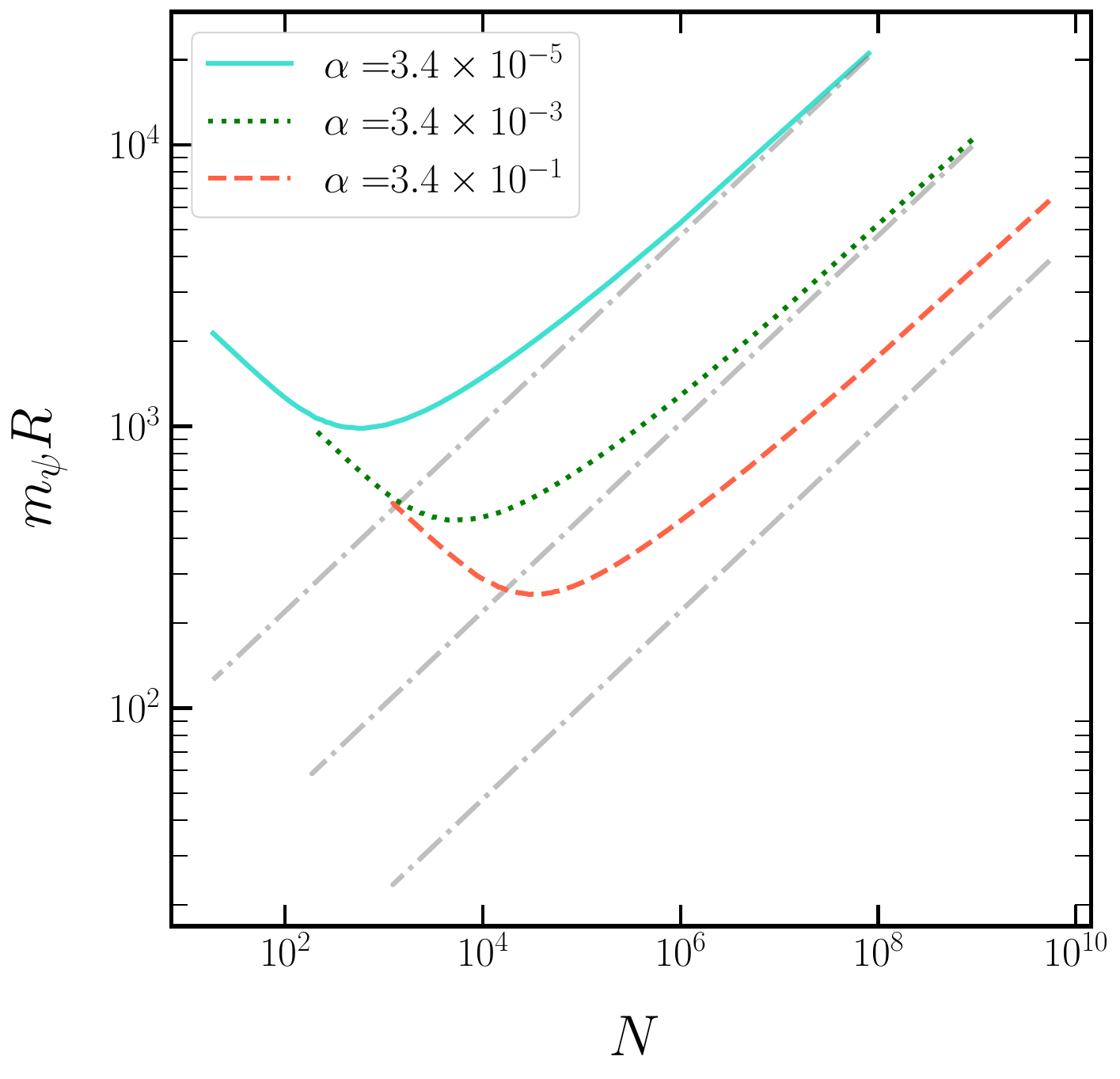}
    \end{subfigure}
    \begin{subfigure}{0.48\textwidth}
        \includegraphics[width=\textwidth]{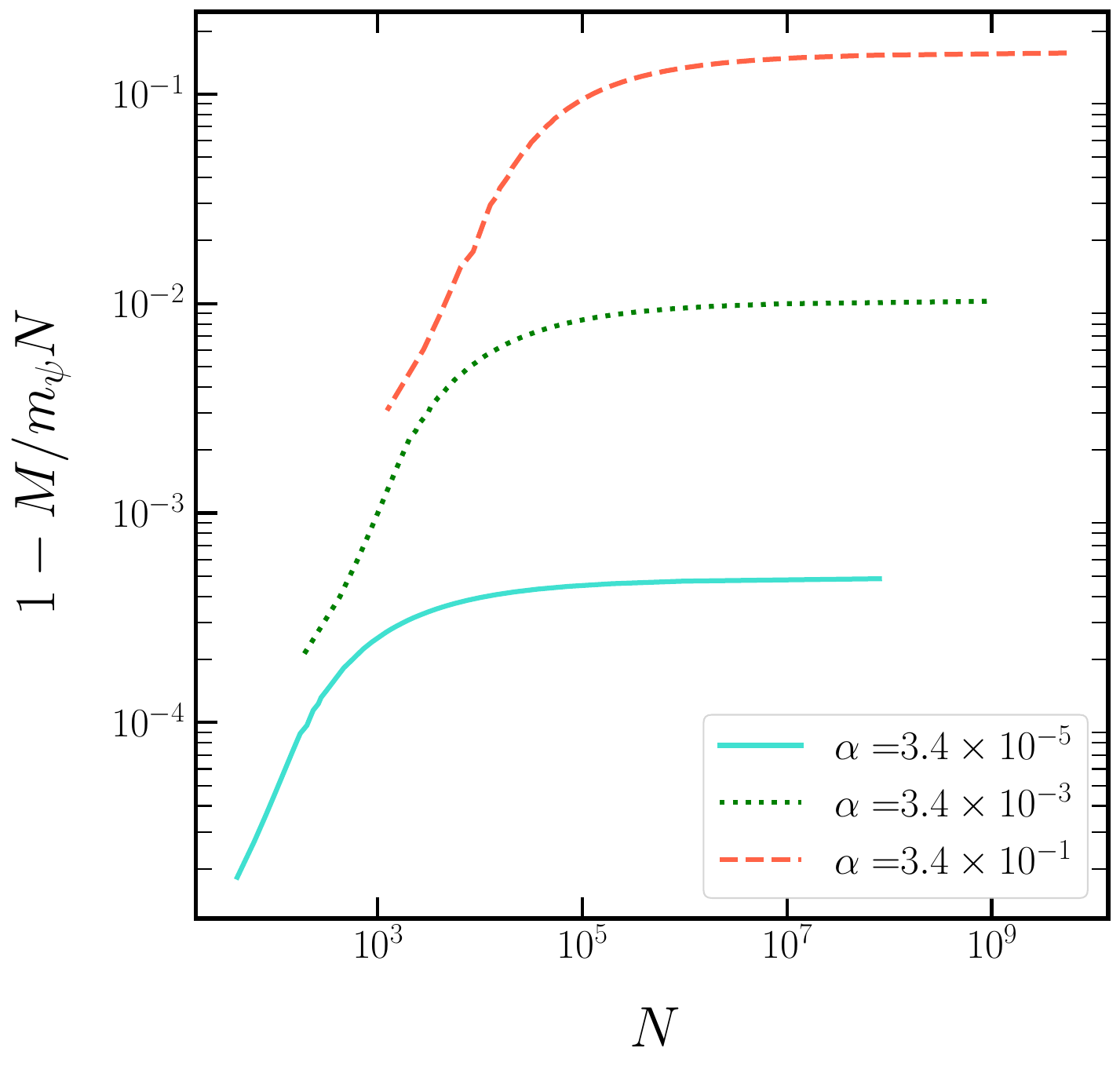}
    \end{subfigure}
    \caption{The scaling of the Fermi ball radius (left panel) and binding energy (right panel) with the number of fermions $N$. The gray dash-dotted lines in the left panel are analytic solutions from \equaref{rmscaling}. We fix $\mpsi/\mphi = 10^6$ in this plot.}
    \label{fig:scaling}
\end{figure*}

\noindent The approximation is accurate as long as $\tilde{\varphi}, \kappa \ll 1$, so we need $y$ to be small enough so that $\alpha \ll 1$. For $\lambda \sim \mathcal{O}(1)$, this merely restricts $y\lesssim 1$; this ensures that Fermi ball -- Fermi ball interactions are strong enough to have non-trivial consequences for mergers, while also guaranteeing rapid Fermi ball formation if they are formed from the same Yukawa force.  The Fermi ball radius and mass can then be calculated as
\begin{equation}
\begin{aligned}
    R &= \frac{4}{5} \left(\frac{9 \pi}{2 g_f}\right)^{1/3} \frac{N^{1/3}}{\alpha^{1/3}\mpsi},\\
    M &= \mu N \simeq \mpsi N.
\end{aligned}
\label{rmscaling}
\end{equation}
We also solve the scalar equation of motion numerically and obtain numerical results for $R$ and $M$. A comparison of our strong coupling branch solutions with the numerical results is demonstrated on the left panel of \figref{fig:scaling}, and on the right panel, we plot the (normalized) binding energy per fermion number \mbox{$1 - M/(\mpsi N)$.} Here, the saturation behavior is apparent from the plateau region of the binding energy. Notably, at small $N$ we observe the characteristic $R\sim N^{-1/3}$, expected for the nugget regime, whereas for larger $N$ the Fermi balls quickly saturate to the analytic solution of \equaref{rmscaling}. As $\alpha$ increases, the analytic solution deviates from the numerical radius, but the $N^{1/3}$ scaling still holds. It is also interesting to note in \figref{fig:scaling} the absence of the sub-saturated $R\sim N^{2/3}$ scaling at intermediate $N$: This is due to the fact that in the small $\mphi$ limit, the mass term decouples from the Fermi ball equation of motion and the effective scalar mass is solely determined by $\sqrt{\lambda} \varphi$ inside of ball. 

We can estimate the fermion number at saturation by comparing the effective scalar mass with the ball radius:
\begin{equation}
    R_{\rm sat} \sim \frac{1}{m_{\varphi, \rm eff.}} \sim \frac{1}{\sqrt{\lambda} \varphi}.
\end{equation}
This gives the critical fermion number at saturation,
\begin{align}
    N_\mathrm{sat.} \sim  \frac{1}{y \sqrt{\lambda}}.
\end{align}

\noindent In simpler terms, the situation is as follows: The inclusion of the $\varphi^4$ term in the potential makes the effective mass of the scalar field inside the Fermi ball significantly larger than without the quartic interaction. Since a Fermi ball becomes saturated when its size is comparable to $1/m_{\varphi,\rm eff.}$, these Fermi balls now reach saturation at significantly {\it smaller} fermion number $N$. This is then the ideal situation for black hole formation---the Fermi ball radius can indeed become smaller than its Schwarzschild horizon, but the Yukawa interaction length is now much larger than the Fermi ball itself, allowing for Fermi ball -- Fermi ball mergers and growth, as discussed in the next section.


\section{Fermi ball -- Fermi ball interactions}\label{sec:interaction}
\noindent In order to understand the Fermi ball -- Fermi ball interactions, we must examine the scalar field solution outside the ball. Due to the lack of a large scalar mass, $\varphi$ decreases more slowly compared to the saturated Fermi ball with $V = 0$, and it satisfies,
\begin{equation}
    \frac{d^2 \varphi}{dr^2} + \frac{2}{r}\frac{d \varphi}{dr} - \lambda \varphi^3 = 0~.
    \label{eq:phiodeoutside}
\end{equation}
The total energy in the two Fermi ball system is given by,
\begin{equation}\label{eq:energy}
    E(r) = -\frac{1}{2} \int y \varphi(r) \braket{\bar{\psi} \psi} d^3 x ~,
\end{equation}
where $\braket{\bar{\psi} \psi}$ can be physically understood as the position-dependent fermion density in the non-relativistic limit. Then the force between two fermi balls can be derived from taking the derivative of the total energy with respect to the separation between the Fermi balls.

\subsection{Single ball solution}
\noindent Deriving the above interaction energy is, however, not a straightforward task. \equaref{eq:phiodeoutside} governs the exterior behavior of the scalar field, which also appears in the context of Fubini instanton \cite{Fubini:1976jm, Linde:1981zj}. However, this nonlinear, second order differential equation is further complicated by the boundary condition at the surface of the ball, which is determined by the interior solution of \equaref{rhoode} with no simple, analytical solution. Even in the single ball case, the solution is elusive. While we could use numerical shooting methods to solve the boundary value problem for $\varphi$ \textit{within} the Fermi ball, the field behavior at longer distances becomes exponentially sensitive to the initial conditions. As a result the solution outside the ball becomes rapidly unreliable---in fact we do see related numerical artifacts in our numerical solution where the scalar field outside of the ball dips below zero.

To make inroads, we first define $n(r) = r \varphi$, so that we now have the marginally simpler equation
\begin{equation}
    \frac{d^2 n}{dr^2} - \lambda \frac{n^3}{r^2} = 0~.
    \label{chargeode}
\end{equation}
The quantity $n(r)$  can be understood as the distance-dependent effective {\it charge} of the Fermi ball, up to a factor of $y$. To see this, consider the field equation without the non-linear coupling:
\begin{equation}
    \frac{d^2 n}{dr^2} = 0.
\end{equation}
The solution for $n$ that matches the boundary condition for $\varphi$ at $r\rightarrow \infty$ and $r \rightarrow R$ is simply a constant; for a single test particle such solution gives an interaction energy
\begin{equation}
    E(r) = -\frac{y n}{r}.
    \label{potential1particle}
\end{equation}
In the expression above it is clear that $n$ corresponds to the effective charge of the ball. In fact, this charge can be determined by the boundary condition
\begin{equation}
    n(R) = \varphi(R) R = \left(\frac{3 y}{4 \pi \lambda}\right)^{1/3} N^{1/3}.
\end{equation}
Note that the unusual factor of $N^{1/3}$ in the effective charge can be understood by rewriting it using the fermion number at saturation $N_{\rm sat} = 4 \pi/(3 y \sqrt{\lambda})$, which yields
\begin{equation}
    n(R) \sim y N_{\rm sat} \frac{R}{R_{\rm sat}},
    \label{eq:maxcharge}
\end{equation}
where we used \equaref{rmscaling}. We therefore estimate the scalar field solution as 
\begin{align}\label{eq:fieldsol}
    \varphi(r) \simeq y~ \frac{N_\mathrm{sat}~R}{R_\mathrm{sat}~r}.
\end{align}
This estimation, however, is only valid near the ball surface. Restoring the nonlinear term introduces additional screening of the charge that also depends on $r$. Nevertheless, we expect such screening to be subdominant compared to the ordinary exponential screening due to the scalar mass for the following reason: the screening length scale from \equaref{chargeode} is $r/(\sqrt{\lambda} n)$, which is enhanced with increasing $r$ and decreasing $n$. This is particularly important when Fermi balls are separated by cosmological distances, i.e.  $r \gg R$. Furthermore, the screening effect should be greatly reduced when the effective charge is close to $N_{\rm sat}$: We can see this by setting 
\begin{equation}
    \frac{n(r)}{r^2 d^2 n/d r^2} = \frac{1}{\lambda n^2} \gtrsim 1.
\end{equation}
Recalling $n(r) = y N$, we find the critical charge to be $N \lesssim N_{\rm sat}$ so that the nonlinear effect becomes negligible. 

Therefore, while we cannot solve the differential equation exactly, we expect that the effective charge will lie in the range
\begin{align}
    N_{\rm sat} \lesssim N_\mathrm{eff.} \lesssim N_{\rm sat} R/ R_{\rm sat}~.
    \label{eq:chargerange}
\end{align}
Interestingly, an extremely similar scalar field solution with nonlinear self-coupling is studied in the case of fifth force screening from chameleon/symmetron fields \cite{Burrage:2018xta, Burrage:2021nys, Tamosiunas:2021kth} and in the long range interaction between Bose-Einstein condensates of dark matter \cite{Berezhiani:2018oxf}. Unfortunately, analytic results can be obtained only when $\lambda \ll 1$, while numerical solutions are similarly plagued by the unstable nature of the ball-shooting process after the source becomes large (saturated in our case). Finding the behavior of $n(r)$ at a full range of $r$ requires a dedicated numerical solver which is well beyond the scope of this work.  In fact, in the context of the chameleon mechanism, the motivation for these fields is to suppress deviations from general relativity \textit{closer} to the sources, so we may have a similar situation here where the screening is relatively larger at close distances (Of course, the scalar field should become exponentially suppressed past the Yukawa interaction length no matter what.)


\subsection{Two ball estimation}

\noindent Given the estimate of the effective charge of a single saturated Fermi ball outlined above, we now discuss the resulting force between \textit{two}, saturated, Fermi balls. The exact solution here is significantly more complicated than in the single case, primarily because one now needs to solve the scalar field equation in {\it three dimensions}, where the boundary conditions are set on the spherical surfaces of the two balls and at infinity.

Solving this exactly is beyond the scope of this paper, but we can make general arguments towards the solution as follows: Let us first assume that the solution is of the form $\varphi = \varphi_1+\varphi_2$, where $\varphi_i$ is the solution to the isolated ball problem. This is of course not exact, since our differential equation is nonlinear and therefore cannot generically be decomposed into the sum of two independent solutions---however, when the balls are relatively distant, the change in scalar field near one ball due to the other is small. This ansatz will be worst in the middle of the two balls, where the contributions to the scalar VEV from each ball is of the same order, but this region does not directly contribute to the interaction energy of the two balls. 

Of course, a large scalar VEV will {\it screen} the distant scalar field close to the ball surface. We then make an additional assumption---the isolated ball solution $\varphi_i$ should drop to zero just {\it within the surface} of the other ball. This is essentially enforcing that only the exterior shell of the Fermi ball contributes to their interaction. We naturally take the width of this shell to be the screening distance $1/m_{\varphi,\rm eff}$. More precisely, the scalar field would be continually screened as it gets closer to the ball surface, and would more smoothly drop to zero in the ball interior. Considering the exterior shell should provide an adequate approximation for this behavior in lieu of an exact solution. 

The number of fermions in the outer shell is given by,
\begin{align}
    N_s \sim \left(\frac{36\pi\lambda}{y}\right)^{1/3}N^{2/3}~.
\end{align}
Inserting the solution of Eq.~(\ref{eq:fieldsol}) into the interaction energy, Eq.~(\ref{eq:energy}), and taking the derivative with respect to the two body separation $r$, we can find the force between two fermi balls. Since we give a range for the effective charge, we also give an upper and lower bound for the force, $F_\mathrm{low}$ and $F_\mathrm{high}$:
\begin{align}
    F_\mathrm{low} &\simeq \left(\frac{243}{16\pi^2}\right)^{1/3}\frac{y^{2/3}\lambda^{-1/6}}{r^2}\left(N_1^{2/3}+N_2^{2/3}\right)~,\nonumber\\
    F_\mathrm{high} &\simeq 3 \frac{y}{r^2}\left(N_2^{1/3}N_1^{2/3} + N_1^{1/3}N_2^{2/3}\right)~,
    \label{eq:forcerange}
\end{align}
where we note that the force has the somewhat unusual property of involving the sum of charges. In \equaref{eq:forcerange}, we substitute the range of effective charge from \equaref{eq:chargerange}, so $F_\mathrm{high}$ remains greater than $F_\mathrm{low}$ as long as the balls are saturated.

One could ask what differs in this procedure from the more standard case of two electrically charged objects, possibly with different charges. One could follow a similar procedure to what outlined above, replacing the exterior scalar potential with the vector potential. However, because the equations of electromagnetism are linear, one would find that the vector potential is proportional to the charge of the ball and so both terms in the sum end up being identical. Here, in contrast, the scalar potential is {\it not proportional to the charge in the ball shell}, but is proportional, rather, to $N^{1/3}$, giving us the asymmetry in the sum shown in the equation above.

Let us now compare the strength of the Fermi-ball interaction to gravity. A simple estimation comparing the strength of gravitational attraction to the higher estimate of the force gives us,
\begin{align}
     \frac{F_\mathrm{high}}{F_\mathrm{grav}} \simeq 0.1~y\left(\frac{10^{10}~\mathrm{g}}{M}\right)\left(\frac{10^5~\mathrm{GeV}}{m_\psi}\right)~.
\end{align}
The Fermi ball mass range relevant to the long range interaction is perhaps surprisingly small---once Fermi balls form, the screening of the Yukawa force is so large, even in our higher estimation, that the force can easily be weaker than gravity for reasonably large Fermi balls. However, the subsequent growth of lighter Fermi balls is still interesting, which could lead to late time black hole formation \cite{Picker:2023ybp, Picker:2023lup, Lu:2022jnp, chakraborty_formation_2022}. It would be prudent to return to this result, however, once we have a better numerical analysis of the Fermi ball-Fermi ball force, because even a small change to the scaling of those relations with respect to $N$ can have a large impact on this conclusion. We hope to investigate better numerical methods in the future.

\section{Fermi ball formation and cosmology}\label{sec:cosmology}
\noindent The model of Fermi balls discussed in \secref{sec:fballs} leads to several different possible realizations in cosmology. Here, we discuss a particularly simple yet generic way of forming Fermi balls in the early Universe, through the same Yukawa interaction that enables rapid structure formation even in the radiation dominated era, as discussed in \refref{Flores:2020drq, Gradwohl:1992ue, Gubser:2004uh, Nusser:2004qu, Amendola:2017xhl, Savastano:2019zpr, Flores:2023zpf, Domenech:2021uyx}. In this section, we first review this Fermi ball formation mechanism, and then discuss the possible constraints from cosmology.

\subsection{Formation mechanism}

\begin{figure*}[ht!]
    \centering
    \includegraphics[width=.8\textwidth]{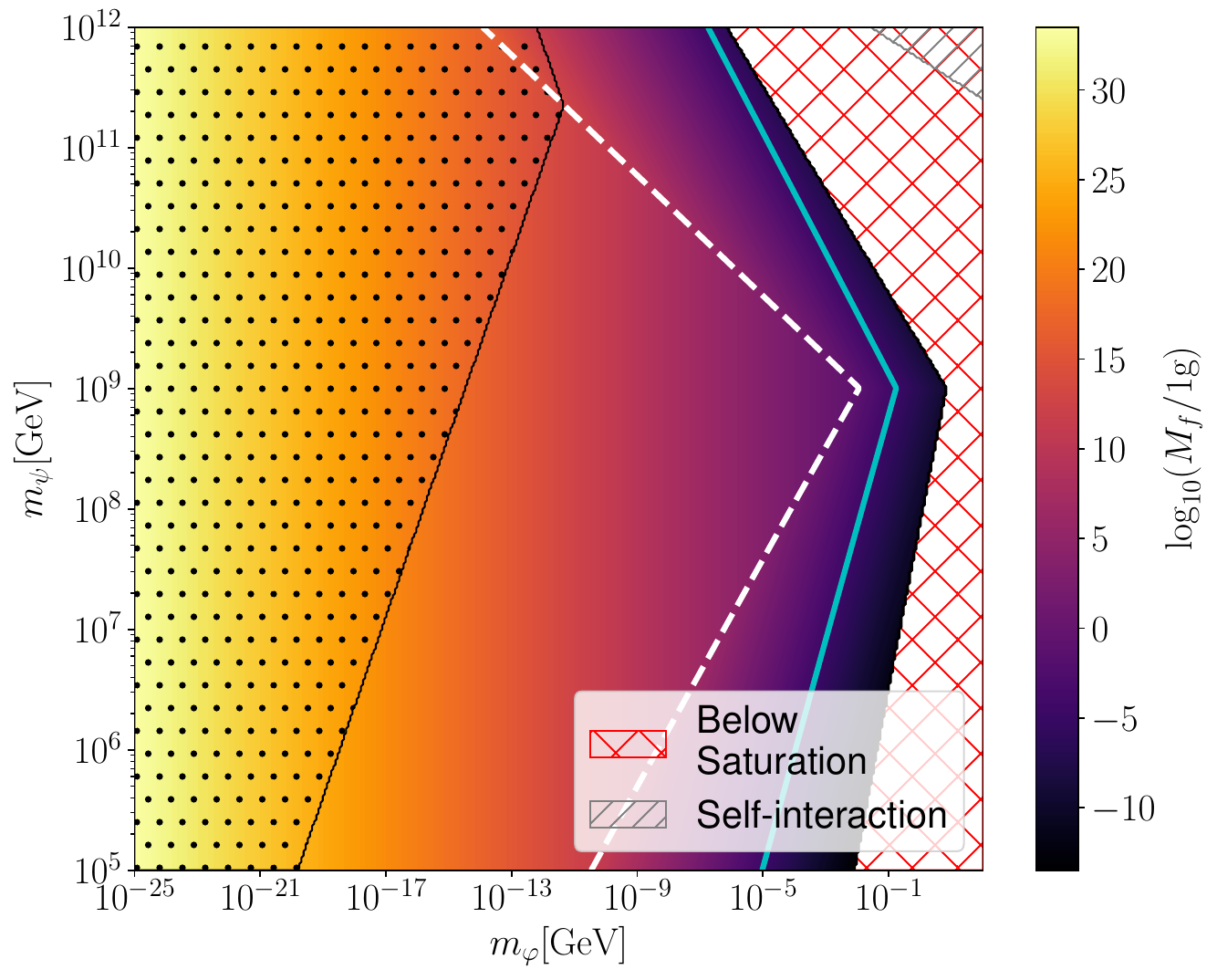}
    \caption{Mass contours for the compact objects formed in the non-instantaneous and instantaneous scenarios (separated by the white dashed line, with non-instaneous to the left) with $f_\mathrm{DM} = 1$. The region covered in black dots indicates immediate black hole formation while the rest of the parameter space gives Fermi balls. The solid cyan line marks Planck mass Fermi balls and the red meshed region corresponds to Fermi balls below saturation. The dark matter self-interaction constraint is marked by the gray shaded region (top right corner). Parameters are chosen to be $y = 5 \times 10^{-2}$, $\lambda = 1 \times 10^{-2}$ and $\eta$ is determined via \equaref{eq:fdm}.}
    \label{fig:parameterspace}
\end{figure*}

\noindent The presence of the Yukawa interaction in the dark sector creates an  attractive force between $\psi$ fermions in addition to gravity. Before compact objects are formed, the interaction between fermions simply follows the non-relativistic Yukawa potential:
\begin{equation}
    V(r) = \frac{y^2}{r} e^{-\mphi r}.
\end{equation}
In general, the interaction strength is stronger than gravity when $\beta \equiv y M_{pl} / \mpsi \gg 1$, where $M_{pl}$ is the Planck mass. Then for a population of out-of-equilibrium and non-relativistic fermions $\psi$, such Yukawa interaction yields rapid growth of the density perturbation $\Delta_k = \left(n_\psi-\bar{n}_\psi\right) / \bar{n}_\psi$ with mode $k \gg \mphi$. In the linear regime, the growth of the density perturbation follows,
\begin{equation}
    \Delta_k(t) =\Delta_k(0) I_0\left(4 \sqrt{p}\left(t / t_{\mathrm{eq}}\right)^{1 / 4}\right)~,
    \label{eq:growth}
\end{equation}
where $I_0$ is the modified Bessel function of the first kind,  $t_{\mathrm{eq}}$ is the time at matter-radiation equality and $p \sim (1 + \beta^2) \gg 1$ \cite{Flores:2020drq}. Contrary to the ordinary structure formation under gravity, \equaref{eq:growth} exhibits exponential growth (even in the radiation dominated era) if $\sqrt{p} (t / t_{\mathrm{eq}})^{1/4} \gg 1$, which happens as long as the temperature is below
\begin{equation}
    T_{\text {form }} \simeq \beta^2 T_{\text {eq }},
\end{equation}
where $T_{\text{eq}}$ is the temperature of the Universe at matter-radiation equality.

Combining this with the condition that $\psi$ falls out of equilibrium with the relativistic, radiation-like background consisting of the light $\varphi$,  and that $\psi$ is non-relativistic at the onset of structure formation, we obtain the temperature of the Universe at which structures begin to grow:
\begin{equation}
    T_g=\min \left(\frac{m_\psi}{3}, \frac{m_\psi}{\ln \left(y^4 m_p / m_\psi\right)}, T_{\text {form }}\right).
\end{equation}
The rapid phase of exponential growth generally results in compact, filamentless halos as demonstrated with simulations in \refref{Domenech:2023afs}. The same Yukawa interaction makes it possible to dissipate energy away from the halo in the form of radiated $\varphi$ particles \cite{Flores:2020drq}. Therefore, the halo does not virialize and  continues to collapse to Fermi balls or black holes, with a timescale typically shorter than the Hubble time. 

The relic density of a population of dark sector fermions can be produced from a baryogenesis-like process \cite{Kaplan:2009ag, Zurek:2013wia, Petraki:2013wwa}  characterized by an asymmetry parameter $\eta$. Here, instead of taking $\eta$ to be related to the baryon asymmetry parameter $\eta_B$, as typical in asymmetric dark matter models, we consider a generic scenario where $\eta$ is a free parameter. The fraction of Fermi balls or black holes participating in the cosmological dark matter is then determined by $\eta$ via,
\begin{align}
    f_\mathrm{DM}=\frac{\Omega_{\rm FB}}{\Omega_{\rm DM}} \simeq 0.23\left(\frac{m_\psi}{10^{10}~GeV}\right)\left(\frac{\eta}{10^{-20}}\right)~.
    \label{eq:fdm}
\end{align}
Following \refref{Flores:2023zpf}, compact object formation can be then divided into two scenarios:\\

\noindent\textbf{(i) Instantaneous formation}: if the maximum structure formation length scale $1/\mphi$ is within the horizon at the epoch of growth $T_g$, due to the exponentially growing mode the entire formation process takes place almost instantaneously (compared to the Hubble time scale), followed by a subsequent cooling and collapse phase. The mass of the final compact object is given by the total $\psi$ mass enclosed within a $1/\mphi$ radius,
\begin{align}
    M &\simeq \frac{\eta m_\psi T_g^3}{\mphi^3} \nonumber\\
    &\simeq  4\times10^{10}~\mathrm{g}~f_\mathrm{DM}\left(\frac{10^{-5}~\GeV}{\mphi}\right)^3\left(\frac{T_g}{10^{10}~\GeV}\right)^3~.
\end{align}
We note the curious situation that $f_\mathrm{DM}$ enters as a parameter in the Fermi ball mass itself via the asymmetry parameter, and the dependence on $m_\psi$ only enters through the growth temperature $T_g$.\\

\noindent\textbf{(ii) Non-instantaneous formation}: if the formation length scale $1/\mphi$ is well beyond the horizon at $T_g$, the $\psi$ fermions within the horizon first cluster into halos. Subsequently, as the horizon expands, Yukawa forces drive the merger of such halos until the horizon is of size $1/\mphi$, which happens at the temperature
\begin{equation}
    T_c \simeq \sqrt{\mphi M_{pl}} .
\end{equation}
The resulting compact object has mass
\begin{align}
    M &\simeq \frac{\eta m_\psi T_c^3}{\mphi^3} \nonumber\\
    &\simeq 1.7\times10^{9}~\mathrm{g}~f_\mathrm{DM}\left(\frac{10^{-10}~\GeV}{\mphi}\right)^{3/2}~,
\end{align}
which now does not depend at all on the fermion mass. 

An additional complication pertains however to the non-instantaneous case. As we demonstrated in \secref{sec:interaction}, the Yukawa interaction between Fermi balls experiences additional screening due to the non-linear scalar self coupling. Therefore, in the less likely scenario that the cooling time scale of the halo is \textit{shorter} than the merger time scale, the subsequent clustering would be driven by the merger of \textit{Fermi balls} instead of \textit{halos}, where the former has potentially drastically lower rates. A detailed analysis of the intermediate merger stage requires better understanding of the Fermi ball -- Fermi ball interaction and halo cooling under Yukawa radiation, which we leave for future work. 


The critical mass of Fermi balls for black hole formation can be estimated by comparing their radius with the corresponding Schwarzschild radius $R_s$. According to \equaref{rmscaling}, saturated Fermi balls that are heavy enough will always collapse to black holes, due to the scaling relation $R_s / R \sim N^{2/3}$. In \figref{fig:parameterspace}, we plot the mass contours of the compact object formed in both instantaneous and non-instantaneous scenarios assuming $f_\mathrm{DM} = 1$, with mass and radius computed from the analytic result of \equaref{rmscaling}. 

Fig.~\ref{fig:parameterspace} explores the $(m_\varphi,m_\psi)$ plane and shows contours of the compact object mass. The region shaded with black dots corresponds to immediate {\it black hole} formation, while the remainder of the parameter space leads to stable Fermi balls. The region to the right, with red diagonal shading, corresponds to Fermi balls below saturation. We also show constraints from dark matter self-interaction ~\cite{Markevitch:2003at,Shen:2022opd} in the top right corner, which we discuss in the subsection below.

We emphasize that in the red meshed region Fermi balls can still be formed, but our analytical results for saturated Fermi balls do not apply anymore. It is also prudent to point out that the formation process outlined here absorbs essentially all of the fermions, i.e. there is a negligible population of free fermions left after  Fermi ball formation. This is a result of the Yukawa force being stronger than gravity, a result first shown theoretically in Ref.~\cite{Flores:2020drq} and then verified in simulations~\cite{Domenech:2023afs}, where it was shown that halos indeed rapidly form without filaments. There may also be some relic density of the $\varphi$ field, although computing its abundance is complicated and intrinsically linked to the formation and interactions of the Fermi balls. A thermal population will slightly modify the effective number of degrees of freedom $N_\mathrm{eff}$, although as pointed out in Ref.~\cite{Flores:2020drq} this may even be desirable for the resolution of the Hubble tension.

\subsection{Self-interacting dark matter constraints}
\noindent Fermi ball -- Fermi ball interactions may in principle conflict with constraints on self-interacting or dissipative dark matter~\cite{Markevitch:2003at,Shen:2022opd}. One of the most relevant constraint on the dark matter momentum transfer cross-section $\sigma$ from the Bullet cluster~\cite{Markevitch:2003at}, giving,
\begin{align}\label{eq:self}
    \frac{\sigma}{m}f_\mathrm{DM} \lesssim 1 cm^2/g \simeq 4.7\times10^3~\GeV^{-3}~,
\end{align}
where we have included the fraction $f_\mathrm{DM}$ of dark matter in Fermi balls in the constraint. We can set a conservative limit on our scenario by taking the cross-section to be $\sigma\simeq\pi \mphi^{-2}$, i.e. the Yukawa length; using the relations above for the Fermi ball masses we  derive the constraint
\begin{align}
    \mphi \lesssim 3.4\times10^{-8}~\GeV\left(\frac{T_\mathrm{g,c}}{1~\GeV}\right)^3~,
\end{align}
which only restricts a small portion of our parameter space, shown in the gray shaded region in the upper, right corner of \figref{fig:parameterspace}. Generally, since the inter-Fermi ball spacing is always larger than the Yukawa length (or else,  structures would have grown further at formation),  the constraints on Fermi balls from their self-interactions are weak. This is a somewhat generic statement---when the interaction range is small compared to the dark matter inter-particle spacing, self interaction constraints are suppressed. Then, for almost all of our parameter space, the Fermi balls act as a relatively standard cold dark matter. 

\section{Discussion and Conclusions}\label{sec:conclusions}
\noindent Fermi balls, as a stable non-topological soliton, can naturally form in the early universe from a relatively simple, interacting dark sector. In this work we derived the Fermi ball scaling relations in the case where the dark scalar field potential has a non-negligible $\lambda\varphi^4$ term. The inclusion of such a term adds a repulsion contact force to the scalar field, modifying the Fermi ball mass and radius scaling relations in critical ways. The Fermi balls reach `saturation' significantly quicker than without the potential, meaning they become relatively `puffier' at smaller masses, when there is still a sizable long range force between Fermi-balls. As these Fermi balls grow, their density increases and they can fall within their own Schwarzschild horizon, thus forming  black holes. 

There are a number of future directions of relevance to our findings, some of which we are already investigating: 

Firstly, one may  include a $\varphi^3$ term in the potential, which would comprise an attractive contact term for the scalar field, further affecting the Fermi ball equations of state and creating a complicated hierarchy depending on which terms in the potential dominate. 

Secondly, our estimate of the Fermi ball -- Fermi ball long range interactions is based on upper and lower bounds on the strength of the force. We intend in subsequent work to apply dedicated numerical methods to better understand the long-range behavior of these balls.

Finally, very tiny Fermi balls can have stronger Yukawa-range interactions with each other throughout the history of the universe and even today. It would certainly be interesting to explore the consequences of such a dark matter self-interaction.

\section*{Acknowledgements}
We thank Derek Inman, Marcos Flores and Chulmoon Yoo for useful discussions. This work was supported in part by the U.S. Department of Energy (DOE) Grants No. DE-SC0009937 (A.K., Y.L., Z.P.) and DE-SC0010107 (S.P.). A.K. was also supported by World Premier International Research Center Initiative (WPI), MEXT, Japan; and by Japan Society for the Promotion of Science (JSPS) KAKENHI Grant No. JP20H05853.

This work made use of N\textsc{um}P\textsc{y}~\cite{numpy2020Natur.585..357H}, S\textsc{ci}P\textsc{y}~\cite{scipy2020NatMe..17..261V}, and M\textsc{atplotlib}~\cite{mpl4160265}.

\bibliography{bib.bib}
\bibliographystyle{bibi}
\end{document}